\documentclass{iaus}
\usepackage{graphics}

  \checkfont{eurm10}
  \iffontfound
    \IfFileExists{upmath.sty}
      {\typeout{^^JFound AMS Euler Roman fonts on the system,
                   using the 'upmath' package.^^J}%
       \usepackage{upmath}}
      {\typeout{^^JFound AMS Euler Roman fonts on the system, but you
                   dont seem to have the}%
       \typeout{'upmath' package installed. iaus.cls can take advantage
                 of these fonts,^^Jif you use 'upmath' package.^^J}%
      }
  \else
  \fi

  \checkfont{msam10}
  \iffontfound
    \IfFileExists{amssymb.sty}
      {\typeout{^^JFound AMS Symbol fonts on the system, using the
                'amssymb' package.^^J}%
       \usepackage{amssymb}%

      }{}
  \fi

  \IfFileExists{amsbsy.sty}
    {\typeout{^^JFound the 'amsbsy' package on the system, using it.^^J}%
     \usepackage{amsbsy}}
    {}

\newsavebox{\astrutbox}
\sbox{\astrutbox}{\rule[-5pt]{0pt}{20pt}}

\def\bh{black hole~}

\def\ers{{\rm erg/sec}}
\def\ms{M_{\odot}}
\def\et{et al.\ }
\def\rev{reverberation~}
\def\vFWHM{\ifmmode v_{\mbox{\tiny FWHM}} \else
            $v_{\mbox{\tiny FWHM}}$\fi}
\def\kms{\ifmmode {\rm km\ s}^{-1} \else km s$^{-1}$\fi}
\def\ers{\ifmmode {\rm erg\ s}^{-1} \else erg s$^{-1}$\fi}
\def\apj{ApJ}
\def\mnras{MNRAS}

\title{The Black Hole- Bulge Relations in Active Galactic Nuclei
}
\author{A. Wandel}
\affiliation{The Hebrew University of Jerusalem, Israel
}

\pubyear{2004}
\volume{222}
\pagerange{1--8}
\date{?? and in revised form ??}
\jname{The Interplay among Black Holes, Stars and ISM \\in Galactic Nuclei}
\editors{Th. Storchi Bergmann, L.C. Ho \& H.R. Schmitt, eds.}
\begin{document}

\maketitle

\begin{abstract}

We show that Massive Black Holes of
AGNs  follow the same approximately linear relation with the luminosity of the host spheroid (bulge) found in normal (quiescent) galaxies,
with the black hole mass being ~0.002 that of the bulge. We also demonstrate that Narrow line AGNs seem to have a significantly lower 
Mbh/Lbulge ratio, which implies a new strong correlation between Mbh/Lbulge and the width of the broad emission lines of AGNs (Wandel 2002).
Combining existing data, new observations and extrapolation schemes based on the Faber-Jackson relation, the relation between the BH mass 
and the stellar velocity dispersion in the host bulge of AGNs is constructed. It too agrees with the relation found in quiescent galaxies.
\end{abstract}

\firstsection 
\section{Introduction}

%
Massive Black Holes (MBHs) detected in the centers of many nearby galaxies 
(Kormendy and Richstone 1995, Kormendy \& Gebhardt (2002))
show an 
approximately linear relation with the luminosity of the host bulge, inferring the black 
hole mass is 0.001-0.002 of the bulge. 
In addition to those MBHs detected by techniques of stellar and gas kinematics,
the masses of about three dozen  MBHs in AGNs have been estimated  by reverberation 
mapping of the broad emission-line region. 
High quality reverberation data and virial BH mass estimates are presently available 
for 20 Seyfert 1 nuclei (Wandel, Peterson and Malkan 1999, hereafter WPM) 
and 17 PG quasars (Kaspi \et 2000), recently reanalysed by Peterson et al. (2004).
The virial  estimate has been shown to be consistent with the real BH mass 
(Peterson \& Wandel 1999; 2000).
Previous work suggested that MBHs of active galactic nuclei 
(Seyfert galaxies and quasars) follow a similar relation (Laor 1998; Ho 1999; Wandel 1999). 
New and updated data for AGN confirm that
AGNs and  quiescent galaxies have the same BH-bulge mass relation
(Wandel 2002; McLure \& Dunlop 2002).
Several authors suggested that Narrow Line Seyfert 1s have lower MBH/bulge ratios
(Wandel 1999;2002; Mathur Kuraskiewicz and Czerny 2001, Gruppe and Mathur 2003)
however Botte \et (2004) find in a different sample that NLS1s have the same BH-bulge relation
as broad line AGNs. 
Ferrarese and Merritt (2000) and Gebhardt \et (2000a) have found that MBH masses of inactive
galaxies are better correlated with the stellar velocity dispersion in the bulge 
than with the bulge luminosity. Apparently this relation holds also for AGNs:
the few Seyfert galaxies with stellar velocity data and reverberation BH mass
estimates seem to be consistent with the BH-velocity dispersion relation of inactive galaxies
(Gebhardt \et 2000b), a conclusion strengthened by observations of the
velocity dispersion in  \rev mapped Seyfert galaxies (Ferrarese \et 2001).

\section{The BH-bulge relation of AGNs}
\begin{figure}
\centering
\resizebox{11.5cm}{!}{\includegraphics{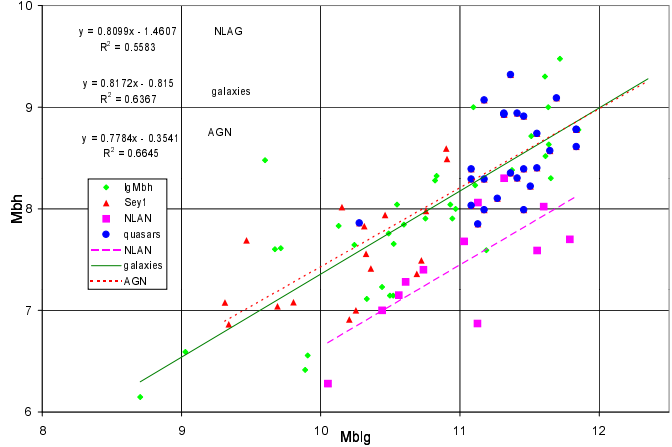} }
  \caption {Black hole mass vs. bulge mass. Green diamonds represent quiescent galaxies,
red triangles and blue circles are broad line Seyferts and quasars, respectively, pink 
squares are narrow line AGNs}
   \label{fig1}
  \end{figure}

We reexamine the BH-bulge mass relation of AGNs.
We combine new and updated data to yield a data base of  55 AGNs (28 quasars,
18 Seyfert 1 galaxies and 9 narrow line Seyfert 1 galaxies and quasars).
Most of these objects have \rev mapping masses and 
almost all of them have
the bulge luminosity measured directly by using  bulge/disk decomposition
(McLure \& Dunlop 2001).
Our results  (figure \ref{fig1}) reconfirm that broad-line
AGNs follow the same BH-bulge relation as ordinary (inactive) 
galaxies, while narrow line AGNs have significantly lower BH/bulge
mass ratios (Wandel 2002).

We find that broad line AGNs have an average black hole/bulge mass fraction of 
$\sim 0.0015$ with a strong correlation ($M_{\rm BH}\propto L_{\rm bulge}^{0.95\pm 0.16}$). 
This BH-bulge relation is consistent with the BH-bulge relation of quiescent galaxies and 
 tighter than previous results (cf. Marconi and Hunt 2003)
 Narrow line AGNs (Narrow Line Seyfert 1s and quasars, defined by permitted lines 
narrower than 2000{\rm km/s} ) appear to have a lower BH/bulge ratio, 
$M_{\rm BH}/M_{\rm bulge}\sim 10^{-4}-10^{-3}$.

The lower BH/bulge ratio of narrow line AGNs may indicate that NLS1s have relatively
 smaller \bh masses. One way to check this possibility is the $\sigma-M_{BH}$ relation,
but it appears that
 the narrow line Seyfert 1 galaxies with measured stellar velocity dispersion have similar 
$\sigma-M_{BH}$
ratios as broad line AGNs and quiescent galaxies (fig 4). 
 
\section{Dependence on the Broad Emission Line Width} 
The lower BH/bulge ratios of narrow-line AGN  seems to be part of a continous trend:
we find that the $M_{\rm BH}/L_{\rm bulge}$ ratio in AGN is strongly correlated with
the emission-line width, 
$M_{\rm BH}/L_{\rm bulge}\sim FWHM(H\beta)^2$
 \begin{figure}
 \centering
\resizebox{11.5cm}{!}{\includegraphics{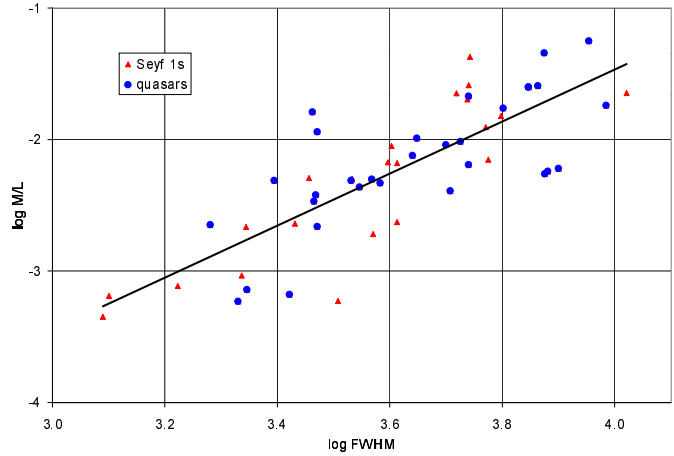}}
  \caption {Black Hole Mass to bulge mass ratio vs. FWHM of H$\beta$. 
Red triangles are broad line Seyferts and blue circles are quasars}  
 \label{fig2}
  \end{figure}

\begin{figure}
 \centering
\resizebox{11.5cm}{!}{\includegraphics{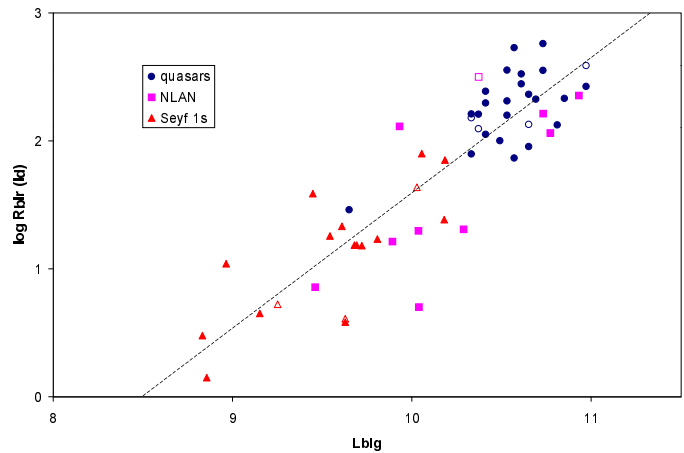}}
  \caption {Size of broad emission-line region (BLR) vs. bulge luminosity. 
Triangles and circles are broad line Seyferts and quasars, respectively, squares are narrow line AGNs}
 \label{fig3}
 \end{figure}

This tight correlation may be related to the virial mass calculation
$M_{\rm BH}=1.46\times 10^5 \ms\left ({R_{blr}\over {\rm lt-days}}\right ) 
\left ({v_{FWHM}\over 10^3\kms }\right )^2$
but given the BH-bulge relation, we expect 
$M_{\rm BH}/L_{\rm bulge}$ to be constant rather than strongly dependent on $FWHM(H\beta)$.
Furthermore, combining the new relation with the virial relation we expect the dependence on the
line width ($FWHM(H\beta)^2$) to cancel and leave a scattered distribution. Rather, we find a new, strong nearly linear relation:
a very tight correlation between bulge luminosity and BLR size (fig.\ref {fig3}).
The best fit of this correlation for all our sample is
$R= 13.5 L_{10}^{1.05}\quad {\rm lt-days} $
with a correlation coefficient of 0.91.
This is a non trivial
correlation, as it relates two independent observables: 
the bulge luminosity, a global galactic property on a scale of kpc,
and the distance of the line-emitting ionized gas
from the central continuum source - the broad  line region (BLR) size on
a scale of few light days.

A possible explanation to this correlation may be as follows: it is known that the BLR radius scales with the AGN luminosity
and the luminosity scales with the accretion rate.  If the accretion rate is determined by the bulge size, the BLR size could
be related to the bulge size.
However, in the BLR-Lbulge relation narrow line AGNs do not seem to have a different scaling than broad line AGNs.
This may indicate that the emission lines
of NLANs appear narrowas a result of an inclination effect, for example,
if the BLR has a flattened geometry and is viewed nearly edge on.


\section{The BH mass - velocity dispersion relation in AGN}

We find that in AGNs with "narrow" boad lines 
 the \bh /bulge ratio seems to be systematically smaller 
than in broad line AGNs and in normal galaxies. Is this the case also for the 
 $M_{\rm BH-\sigma }$ relation?
The bulge velocity dispersion is measured for only 11 Seyfert galaxies,
of which four (Mrk 110, Mrk 590, 3C120 and NGC 4051) can be classified as narrow line Seyfert 1s .
\begin{figure}
\centering
\resizebox{11.5cm}{!}{\includegraphics{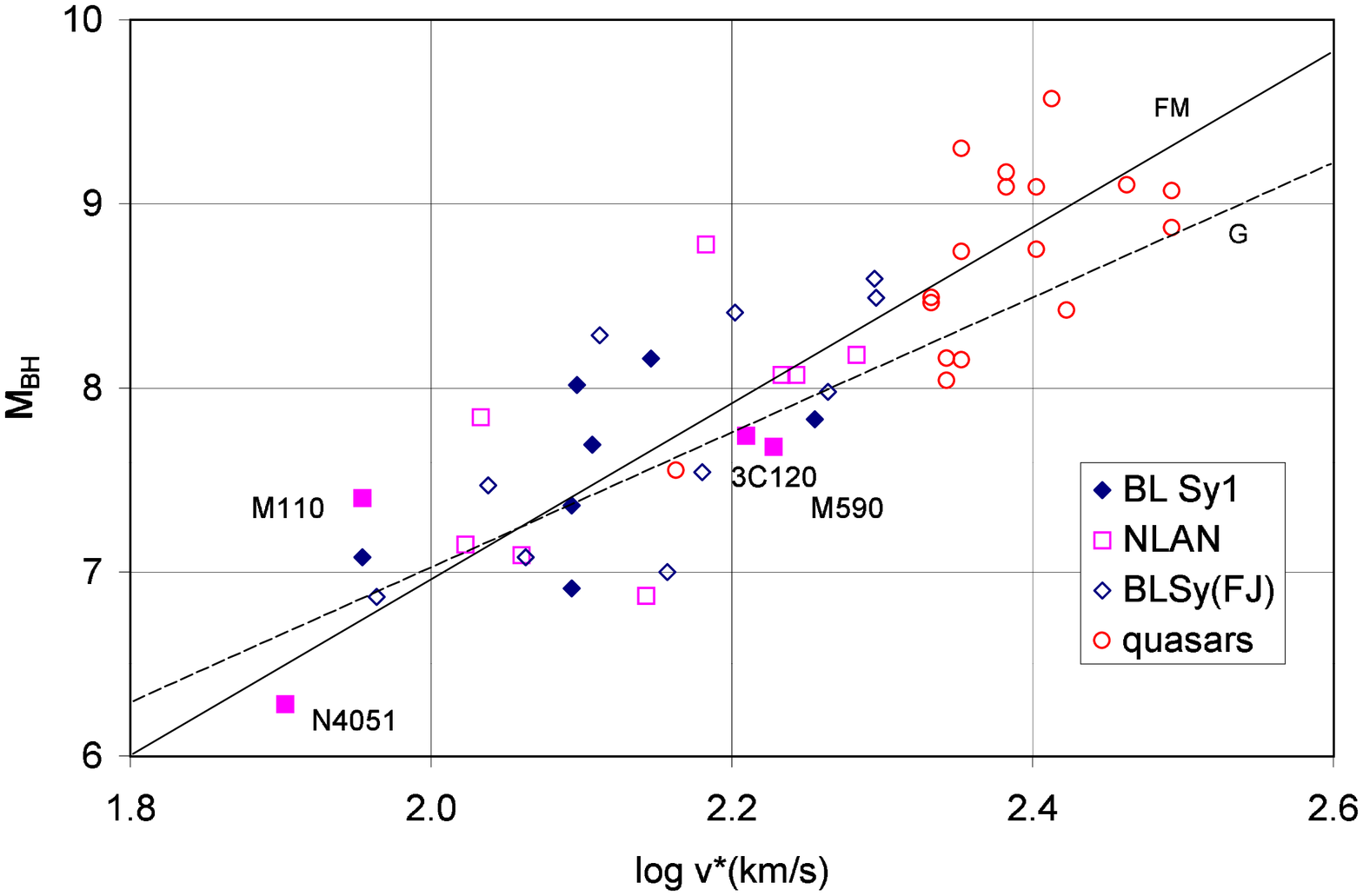}}
\caption{Black hole mass of AGNs plotted against the stellar velocity dispersion.
Blue triangles are broad line Seyferts,  pink 
squares denote NLS1s and red 
circles denote quasars.
Solid symbols denote Seyferts with measured $\sigma$, open ones denote Seyferts 
for which $\sigma$ has been estimated from the Faber-Jackson relation (see text).
The dashed and solid lines show the $\sigma-M_{\rm BH}$ relation of galaxies 
(Gebhardt  \et 2000a and Ferrarese \& Merritt 2000, respectively). 
}
\label{fig4}
\end{figure}

In figure \ref{fig4} the BH mass is plotted against the stellar velocity dispersion. 
We see that broad line AGNs are 
consistent with the $M_{\rm BH}-\sigma$ relation of inactive galaxies 
($M_{\rm BH}\propto\sigma^\alpha$, with  
$\alpha=3.5-5$; Gebhardt \et 2000a; 
Merritt and Ferrarese 2001a) 

The narrow line Seyfert 1s with measured $\sigma$ 
denoted by solid pink squares in figure \ref{fig4}) seem to follow the same 
$M_{\rm BH}-\sigma$ relation of quiescent galaxies and broad line Seyferts. 

For AGNs without a direct velocity dispersion measurement 
 the relation between the narrow line width and the velocity
dispersion (Nelson 2000;Botte \et ) may be used to estimate the velocity dispersion in the bulge.
Interestingly, a tight linear relation seems to exist between virial mass given by the narrow  [OIII] line 
and the BH mass in Seyfert galaxies (Wandel and Mushotzky 1986).

Here we  use a different method to estimate the velocity dispersion - the 
Faber-Jackson (F-J) relation, which is applicable for AGNs with measured or estimated bulge luminosity. 
 The "standard" F-J relation is 
$L= L_o  \sigma_2^4$,
where $\sigma_2=\sigma/100\kms$ and $L_o$ is a luminosity coefficient determined
by a linear fit with a slope of 4 to those Seyfert Galaxies 
in our sample which have measured stellar velocity dispersion,
which gives $L_o=1.6\times10^9\ers$.
Note that most narrow line AGNs added in this manner (open pink squares in fig. \ref{fig4})
do not have lower BH masses than the value expected from the $\sigma -M_{\rm BH}$ relation

\section{Conclusion}

We show that narrow-line AGNs have lower $M_{\rm BH}/L_{bulge}$ ratio than  broad line AGNs,
and find a strong correlation between this ratio and the broad emission-line width.
However, in the $M_{bh}/\sigma^4$ relation narrow line AGNs seem to be similar to broad line AGNs.

\begin{acknowledgments}
This work was supported
by BSF under grant number 1999336.
\end{acknowledgments}

\end{document}